\documentclass{jaa}

\usepackage{natbib}
\usepackage{graphicx}	
\usepackage{url}
\usepackage{hyperref}
\usepackage{siunitx}

\newcommand\hst{{\it HST}}
\newcommand\chandra{{\it Chandra}}

\newcommand\astrosat{{\it AstroSat}}

\newcommand\ks{{\rm~ks}}

\newcommand\kev{{\rm~keV}}

\begin{document}\sloppy

\title{A pair of UV nuclei or a compact star forming region near the active nucleus in Mrk~766?}


\author{P. P. Deka\textsuperscript{1}, G. C. Dewangan\textsuperscript{1}, K. P. Singh\textsuperscript{2} and J. Postma\textsuperscript{3}}
\affilOne{\textsuperscript{1}Inter-University Centre for Astronomy and Astrophysics (IUCAA), SPPU Campus, Pune, India,\\}
\affilTwo{\textsuperscript{2}Indian Institute of Science Education and Research Mohali, Knowledge City, Sector 81,
Manauli P.O., SAS Nagar, 140306, Punjab, India, \\}
\affilThree{\textsuperscript{3}Department of Physics \& Astronomy, University of Calgary, 2500 University Dr. NW, Calgary, AB T2N 1N4, Canada.
}


\twocolumn[{

\maketitle

\corres{gulabd@iucaa.in}

\msinfo{31 October 2020}{16 Dec 2020}

\begin{abstract}
We report the discovery of a bright, compact ultraviolet source at a projected separation of 1.1~kpc from the known active galactic nucleus (AGN) in Mrk~766 based on \astrosat{}/UVIT observations. We perform radial profile analysis and derive the UV flux almost free from the nearby contaminating sources. The new source is about 2.5 and 5.6 times fainter than the AGN in the far and near UV bands. The two sources appear as a pair of nuclei in Mrk~766. We investigate the nature of the new source based on the UV flux ratio,  X-ray and optical emission. The new source is highly unlikely to be another accreting supermassive black hole in Mrk~766 as it lacks X-ray emission. We find that the UV/Optical flux of the new source measured at four different bands closely follow the shape of the template spectrum of starburst galaxies. This strongly suggests that the new source is a compact star forming region.
\end{abstract}

\keywords{galaxies---active galactic nuclei---star formation.}

}]


\doinum{12.3456/s78910-011-012-3}
\artcitid{\#\#\#\#}
\volnum{000}
\year{0000}
\pgrange{1--}
\setcounter{page}{1}
\lp{1}

\section{Introduction}
Presence of dual or multiple compact and luminous sources in the central regions of galaxies are rare. Mergers of galaxies can naturally lead to the formation of dual and/or multiple compact sources such as those observed in Ultraluminous infrared galaxies. Simulations of mergers of gas-rich disk galaxies show that very massive, compact and highly luminous star clusters can form from the strongly disturbed gas disks.  Consist of young stars, these clusters  appear as several bright cores in the central kilo-parsec region of galaxies \citep{Matsui}. 

\par 
Another popular interpretation of dual compact sources is the presence of double accreting supermassive black holes in the central regions.
It is now widely accepted that almost every galaxy has a supermassive black hole (SMBH) at its center whose mass can be estimated through various techniques like central stellar velocity dispersion observations, reverberation mapping observations, etc. \citep{c2}. During the merging of galaxies, their respective SMBHs are also expected to come closer due to gravitational attraction and finally coalesce. The whole process can be classified into three phases \citep{c1}: 1) After the two galaxies merge, the two SMBHs move towards the center of the newly formed galaxy and form a binary pair by losing their angular momentum through dynamical friction. 2) The orbit is further hardened by slingshot ejection of stars whose orbits crosses the binary by the process of three body interaction and thereby moving the two SMBHs closer to each other. 3) When the separation between the two SMBHs is small enough such that emission of gravitational waves can dominate the other forms of energy loss, the two black holes merge. There are many key questions to answer, e.g., till when the two SMBHs retain their individual accretion disks and when they start sharing a common accretion disk, at what rate the accretion takes place and the rate of growth of the individual SMBHs etc. Also, after merging, the anisotropically emitted gravitational wave gives a kick velocity to the final merged black hole due to which it gets ejected from the center \citep{c4}. For sufficient kick velocities, the merged black hole can even get completely ejected from the host galaxy, though its probability is very small \citep{c4}. 

\par


Observing multiple luminous, compact sources in the nuclear regions of galaxies and finding their nature is crucial to understand galaxy evolution and mergers of supermassive black holes.  The exceptional high spatial  resolution of the \chandra{} X-ray telescope and the Hubble Space Telescope (\hst{}) have led to the first discovery of dual nuclei in galaxies~\citep{Komossa_2002,Junkkarinen_2001,2004ApJ...600..634B}. A number of other techniques have also been used (see~\cite{c3} for a review). Here we used high resolution of \astrosat{}/UVIT and discovered a compact, bright UV source near the well known active nucleus in Mrk~766. In Section~\ref{sec:obs}, we present \astrosat{} observations and the reduction of UVIT data. In Section~\ref{sec:ana}, we perform spatial analysis of UVIT images and investigate the nature of the new UV source in Section 4, and summarize our findings in Section 5.


\section{\astrosat{}/UVIT observations and the data reduction}\label{sec:obs}
We observed Mrk~766 with \astrosat{} as a part of the SXT Guaranteed programme during 4-6 February 2017 with  the SXT as the primary instrument for an exposure time of $50\ks$.  Here we present the UVIT data only. We used the
 broadband filters FUV/BaF2 (F154W) and NUV/Silica (N242W) and acquired photon counting data. We obtained the Level1
data from the \astrosat{} data
archive \footnote{\url{https://astrobrowse.issdc.gov.in/astro_archive/archive/Home.jsp}}
and we processed them using the UVIT pipeline
CCDLAB \citep{2017PASP..129k5002P}. We generated cleaned images for each
orbit, aligned them and created merged  image for each filter. This resulted in  net exposure time of 33.4~ks
(NUV/Silica) and 27~ks (FUV/BaF2).  We derived the astrometric solution
transforming the image coordinates to the world coordinates using the
astrometry.net package ~\citep{2010AJ....139.1782L}. We show the NUV and FUV images of Mrk~766 in Figure~1. In order to show the relative intensities of the two UV sources, we created 2D histogram views of the NUV image in two different scales, namely linear and square root. We show the histograms in Figure~\ref{fig:surface}.

    

\begin{figure*}
    \centering
    \includegraphics[width=16cm]{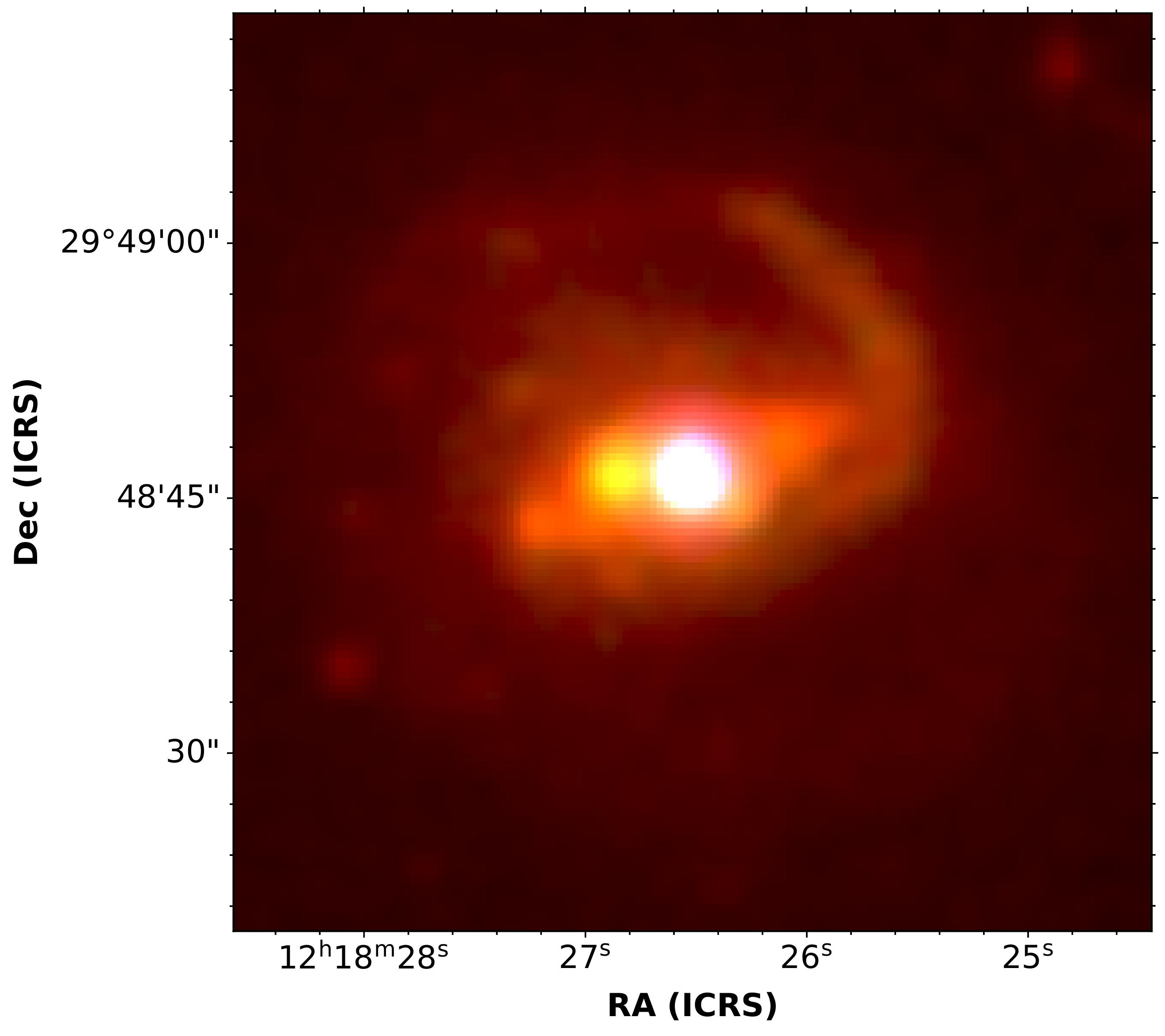}
    \caption{Composite three color NUV/Silica (red), FUV/BaF2 (green) and Chandra/HETG X-ray (blue) image of Mrk~766.}
    \label{fig:col_img}
\end{figure*}

\begin{figure*}
    \centering
     \includegraphics[width=10cm]{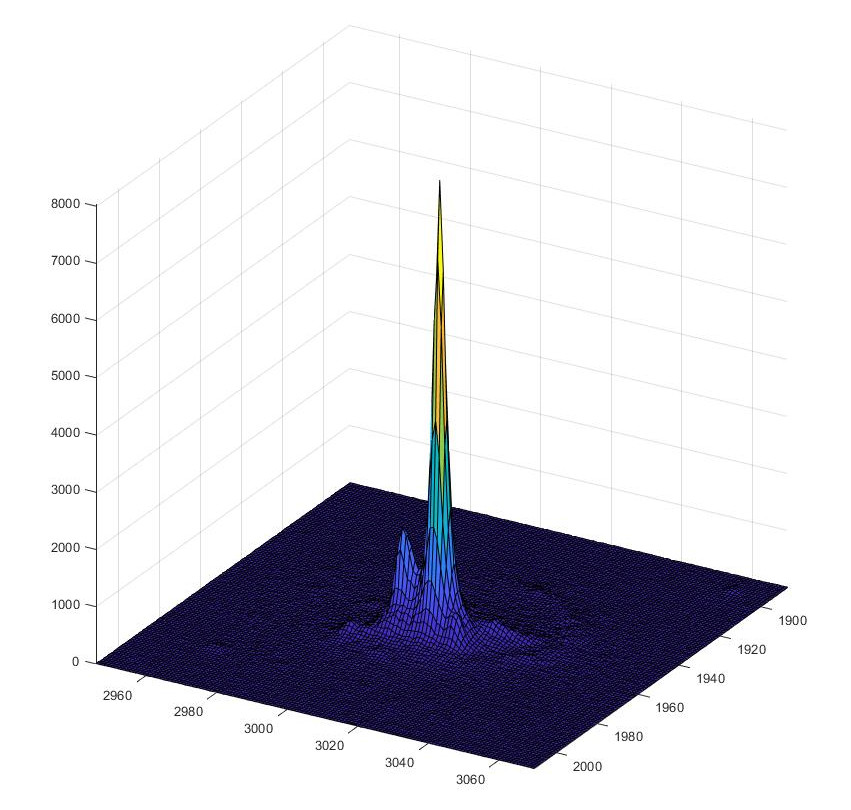}
    \includegraphics[width=7cm]{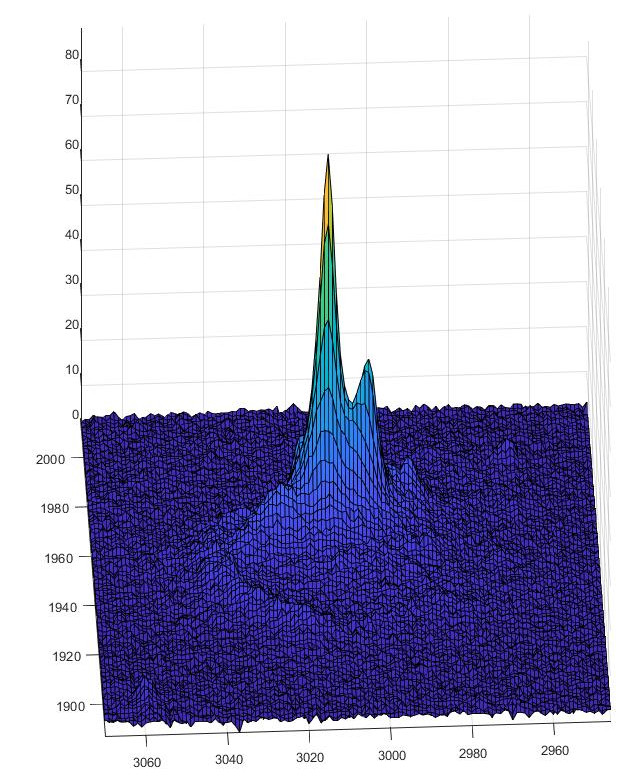}

    \caption{Surface plot of NUV emission from Mrk~766 on linear (left panel) and square root (right panel) scales showing complex spatial structure.}
    \label{fig:surface}
\end{figure*}

To identify the position of known active galactic nucleus in Mrk~766 and to look for X-ray counter part of the new UV source, we generated a composite three-color image using FUV/BaF2, NUV/Silica and \chandra{} X-ray images. We obtained the processed and cleaned  \chandra{}/ACIS event data (obsID:1597) from  the  HEASARC archive\footnote{\url{https://heasarc.gsfc.nasa.gov/cgi-bin/W3Browse/w3browse.pl}}.  The composite three-color image of Mrk~766 is shown in Figure~\ref{fig:col_img}. Clearly the central bright UV source with strong X-ray emission is the known AGN. We refer the AGN as the primary source and the nearby UV source as the secondary source. We did not find significant X-ray emission at the position of the secondary UV source. 

We measured the positions of the primary source (AGN) and the secondary source in both the FUV and NUV images using the centroiding algorithm available with the SAOImage/DS9 tool. We list the positions in Table~\ref{tab:coord}.

\begin{table*}
    \centering
    \caption{Positions of the primary and secondary UV sources}
    \begin{tabular}{ccccc} \hline
      Source   &  \multicolumn{2}{c}{FUV} & \multicolumn{2}{c}{NUV} \\
               & $\alpha$(J2000) & $\delta$(J2000) & $\alpha$(J2000) & $\delta$(J2000)  \\ \hline
   AGN      &  12h18m26.5s & +29d48m46.1s &  12h18m26.5s &  $+$29d48m46.3s\\
   Secondary &  12h18m26.8s & +29d48m46.4s  & 12h18m26.8s & $+$29d48m46.6s \\ \hline
    \end{tabular}
    \label{tab:coord}
\end{table*}


\section{Spatial analysis and measurement of UV flux} \label{sec:ana}
The separation between the AGN and secondary source is $4.211$ arcsec. Though the peak positions of the two sources are well separated, the wings of the two PSFs overlap. Moreover, the two sources are located in the central regions where the diffuse emission from the galaxy is strong. 
In addition, there are other galaxy features such as the bar and star forming clumps possibly associated with spiral arms in the central regions. The small separation and a number of features complicate measurement of flux from the two sources. Here we use a simplistic approach to separate the emission from the two sources. We extract radial profiles and fit with the PSFs of point sources and profile of the diffuse emission and background level. Given the complexity of the spatial structure, our method will be approximate. Unlike optical images, UV images of galaxies are not smooth due to the presence of star forming clumps and possible non-uniformity of internal reddening and even the 2D profile fitting  using tools such as the GALFIT may not yield accurate results. We postpone such a detailed analysis to a future paper. 

Our analysis consists of the following steps in sequence. First  we extract radial profiles centered on one of the two sources in each image. Thus we generate four radial profiles for  the two sources in the NUV and FUV images. We then fit  the radial profiles to obtain the count rates of the AGN and the secondary source in the NUV and FUV bands. If we take the ratio of the NUV and FUV count rates for each source, this will represent the slope of the spectrum of the source. In the final step, we compare these ratios with the slopes of spectrum of known sources such as  quasars and starburst galaxies. Below, we describe our analysis in these steps.

\subsection{Radial profile fitting and count-rate ratio analysis} 

The radial profiles in the FUV and NUV bands centered on each of the two sources were derived using the image display and astronomical data visualization tool SAOImage/DS9. In each case, the source at the center of the radial profile was fitted with a 1D Moffat function (PSF model for UVIT) and the off-centered source was fitted with a 1D Gaussian. The contribution from the galaxy in the form of diffuse emission was fitted with an 1D exponential function and finally the overall constant background was fitted with a constant 1D function. The fitting was performed using \textbf{Sherpa} which is inside Chandra's data reduction and fitting package \textbf{CIAO}. Before going into the results of the fitting process, we would like to state here that due to various components present and resolved in the UVIT images of the galaxy (e.g., the bar in the central region, the extended spiral arm etc.), our simple models for fitting the radial profiles didn't prove  to be sufficient. Consequently, to get acceptable values of the fit statistics, we had to add systematic errors to our data, which resulted in increased error bars in the best-fit parameters. 
Below we give the form of the different profile functions used to fit the components of the radial profile.\par
\begin{equation}
\begin{split}
    &\text{Moffat:}\  m(x)=A\Big[1+\Big(\frac{x-x_0}{\gamma}\Big)^2\Big]^{-\beta}\\ 
    &\text{Gaussian: }\ g(x)= A  \exp \Big[-4 \ln{2}  \frac{(x - x_0)^2 }{ \text{FWHM}^2}\Big]\\
    & \text{Exponential: } e(x)=A  \exp(\text{coeff}  (x - x_0))\\
    & \text{Constant: } c(x)=c_0\\
\end{split}
\end{equation}
In what follows, m1 would indicate a Moffat function, g1 or g2 would indicate Gaussian profiles, e1 would indicate an exponential profile and c1 would indicate a constant background profile.
\subsubsection{FUV radial profile analysis for the primary source}
\begin{figure}[h]
    \centering
    \includegraphics[scale=0.5]{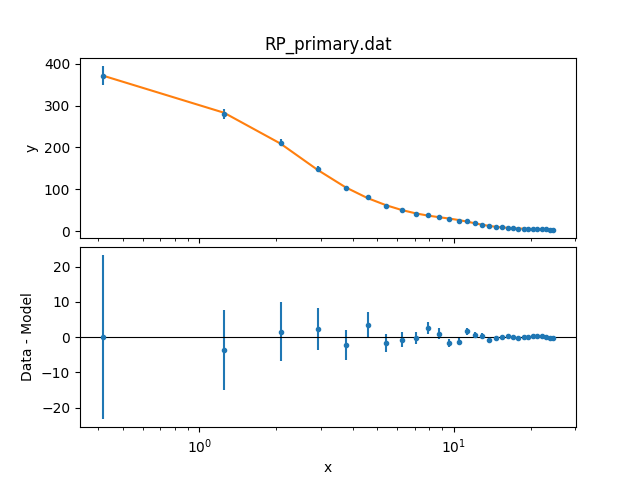}
    \caption{FUV radial profile of the primary source, the best-fitting model and residuals.}
    \label{fig_111}
\end{figure}

For fitting the primary source, we fixed the $\beta$ and $\gamma$ parameters of the Moffat function (m1) by fitting a radial profile extracted from a field star (PSF modelling). The remaining features were fitted with components as described above. Table \ref{tab_0} lists the fitted parameters along with their 89.041$\%$ confidence intervals. Also, we had to add 3.5$\%$ systematic error to get a good $\chi^2$/dof = 21.2/20. Figure~\ref{fig_111} shows the model fitted data along with its residual. We integrated the fitted Moffat function from the position of the peak of the Moffat function to a radius of 25 pixels ($\sim10$ arcsec) which encompasses greater than 95\% of the energy \citep{Tandon_2020} and divide by the exposure time to get the number of counts per second (CPS) from the primary (see Table~\ref{tab_0}).

\begin{table}[htb]
\tabularfont
\caption{Best-fit parameters from the FUV radial profile analysis for the primary  source}
\centering
\begin{tabular}{ lcc}
\hline
Param &Type& Best Value  \\
\hline
m1.A &  thawed & 187.5 $\pm$ 27.3  \\
m1.$x_0$   & frozen    &   0.6 \\
m1.$\gamma$ & frozen  &   2.6 \\
m1.$\beta $ & frozen & 1.9 \\
c1.$c_0$ & thawed  &   2.7$\pm$0.3  \\
g1.$x_0$ & thawed & 10.1 $\pm$ 0.7 \\
g1.A & thawed & 6.4 $\pm$ 1.5 \\
g1.fwhm &thawed & 4.6 $\pm$ 1.2 \\
e1.$x_0$ &thawed & 0.6 $\pm$ 195.0\\
e1.A & thawed &130 $\pm$ 5338 \\
e1.coeff &thawed & $-0.21\pm 0.01$\\
g2.A & thawed & 663 $\pm$ 6169\\
g2.fwhm & thawed & 0.2 + 0.3\\
g2.$x_0$ & linked (m1.$x_0$)& 0.6\\ \hline
CPS &  & $0.23\pm0.03$\\
\hline
\end{tabular}
\label{tab_0}
\end{table}

\subsubsection{FUV radial profile analysis for the secondary source} 
\begin{figure}[h]
    \centering
    \includegraphics[scale=0.5]{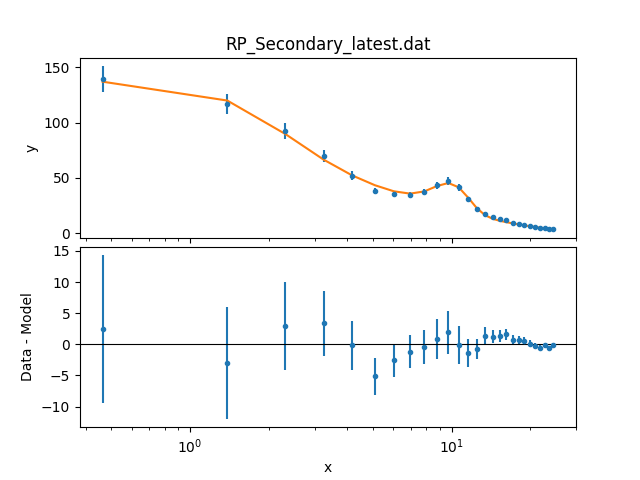}
    \caption{FUV radial profile of the secondary source, the best-fitting model and the residuals.}
    \label{fig_1}
\end{figure}

Again for fitting the secondary source, we fixed the $\beta$ and $\gamma$ parameters of the Moffat function (m1) from the stellar profile. Table~\ref{tab_1} shows the values of the fitted parameters along with their 89.041$\%$ confidence intervals. We had to add 7$\%$ systematic error to get a reasonable value of reduced $\chi^2=1.1$ for 20 dof.  Figure~\ref{fig_1} shows the model fitted data along with the residuals. We integrated the fitted Moffat function from the position of the peak of the Moffat function to a radius of 25 pixels and divide by the exposure time and obtained the count rate of $0.09\pm 0.02$ counts~s$^{-1}$ for the secondary.


\begin{table}[htb]
\tabularfont
\caption{Best-fit parameters from the FUV radial profile analysis for the secondary source}
\centering
\begin{tabular}{ lcc}
\hline
Param &Type& Best Value  \\
\hline
m1.A &  thawed & 72.6 $\pm$ 16.1 \\
m1.$x_0$   & frozen  &   0.6\\
m1.$\gamma$ & frozen  &   2.6 \\
m1.$\beta $ & frozen & 1.9 \\
c1.$c_0$ & thawed  &  0 $\pm$ 1 \\
g1.$x_0$ & thawed & 10.0 $\pm$0.2\\
g1.A & thawed & 23.6 $\pm$3.6 \\
g1.fwhm &thawed & 3.7 $\pm$0.6 \\
e1.$x_0$ &linked (g1.$x_0$)&  10.0 \\
e1.A & thawed &20.9 $\pm$2.1\\
e1.coeff &thawed & $-0.12\pm$ 0.01 \\ \hline
CPS & & $0.09\pm0.02$ \\ 
\hline
\end{tabular}
\label{tab_1}
\end{table}

\subsubsection{NUV radial profile analysis for the primary source}
\begin{figure}[h]
    \centering
    \includegraphics[scale=0.12]{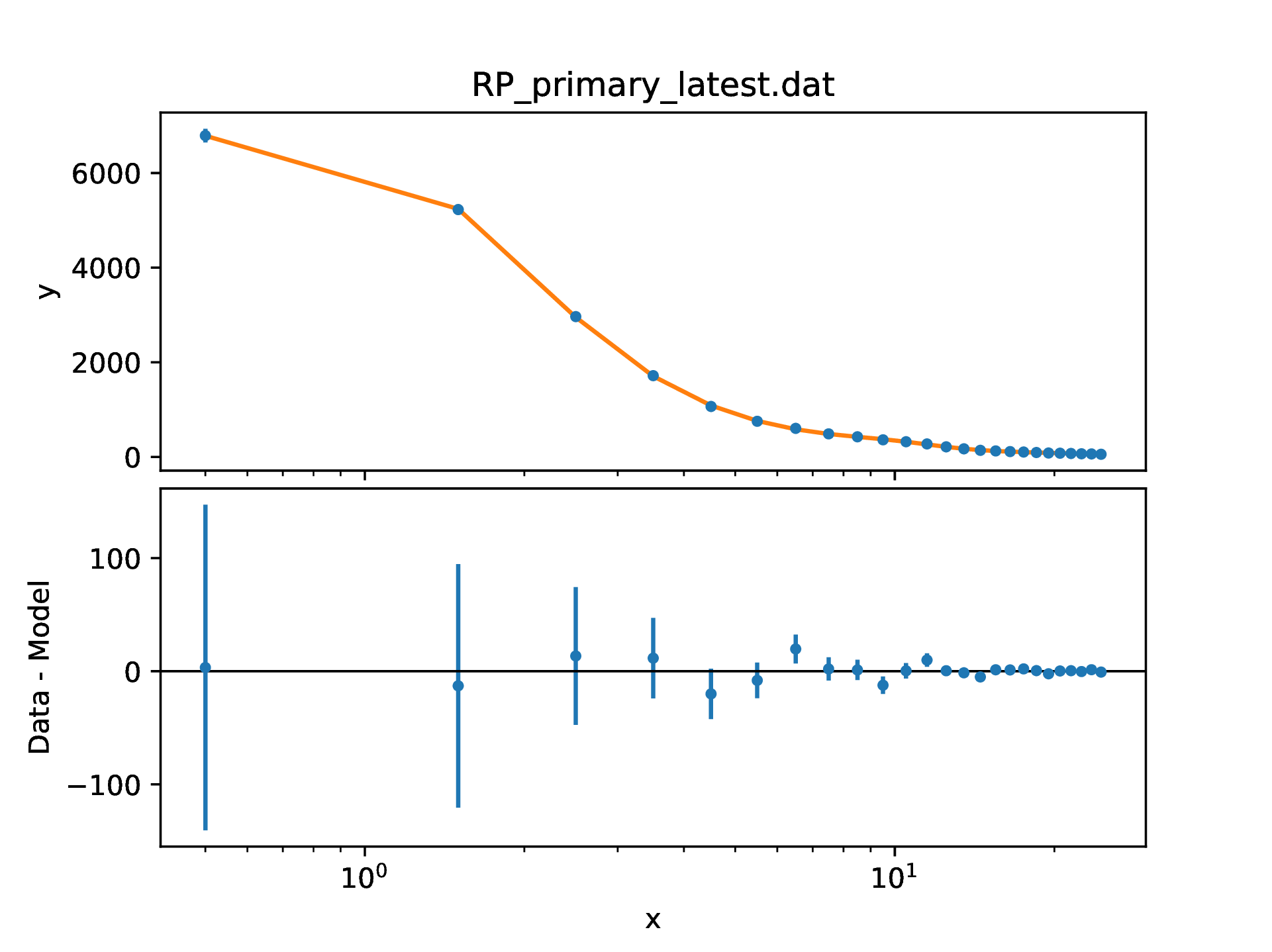}
    \caption{NUV radial profile of the primary source, the best-fitting model and the residuals.}
    \label{fig_2}
\end{figure}

Similar procedures as described above for the FUV band were followed for fitting the NUV radial profiles for the primary as well as the secondary. Table~\ref{tab_2} lists the best-fit parameters along with their 89.041$\%$ confidence intervals. We had to add 2$\%$ systematic error to get a reasonable value of $\chi^2$/dof=25.7/24. In this case, we didn't freeze the Moffat parameters as the bright AGN was good enough to estimate the $\gamma$ and $\beta$ parameters.
\begin{table}[htb]
\tabularfont
\caption{Best-fit  parameters from the NUV radial profile analysis for the primary source}
\centering
\begin{tabular}{ lcc }
\hline
Param &Type& Best Value \\
\hline
m1.A  & thawed   &     5353.3 $\pm$622.8 \\
m1.$x_0$   & thawed     &  0.5 $\pm$0.2  \\
m1.$\gamma$ & thawed  & 2.6 $\pm$ 0.9 \\
m1.$\beta $ & thawed & 2.2  $\pm$ 1.1\\
c1.$c_0$ & thawed  &  44.3 $\pm$ 5.6\\
g1.A & thawed & 58.4 $\pm$ 17.4\\
g1.fwhm &thawed & 3.9 $\pm$  1.1 \\
g1.$x_0$ & thawed & 10.2 $\pm$ 0.6\\
e1.$x_0$ &linked (m1.$x_0$)& 0.5 \\
e1.A & thawed &1384.7 $\pm$ 585.4 \\
e1.coeff &thawed & $-0.19 \pm$ 0.03 \\  \hline
CPS & & $3.9\pm0.4$ \\ 
\hline
\end{tabular}
\label{tab_2}
\end{table}
Figure~\ref{fig_2} shows the model fitted data along with its residual.
We derived an NUV count rate of $3.9 \pm 0.4$ for the primary source using a circular region with a radius of 25 pixels. 

\subsubsection{NUV radial profile analysis for the secondary source}
\begin{figure}[h]
    \centering
    \includegraphics[scale=0.12]{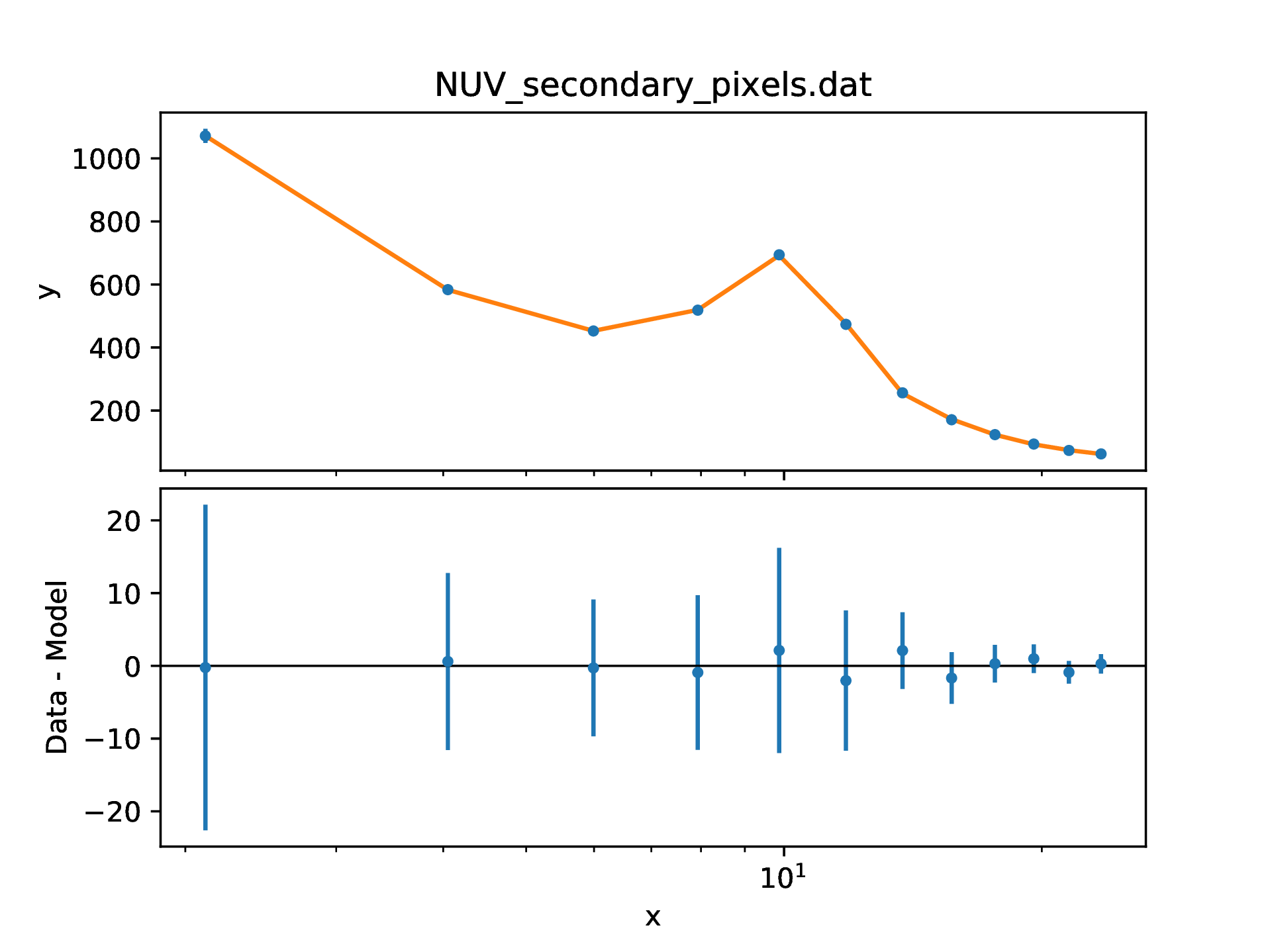}
    \caption{NUV radial profile of the secondary source, the best-fitting model and the residuals.}
    \label{fig_3}
\end{figure}

In this case, we had to fit an additional Gaussian (g2) to account for the AGN. Table~\ref{tab_3} below shows the values of the fitted parameters along with their 89.041$\%$ confidence intervals. Again, we had to add 2$\%$ systematic error to get a reasonable value of reduced $\chi^2=1.09$. In this case, Moffat parameters were fixed based on the stellar radial profile analysis. As before, we derived an NUV count rate of $0.7\pm0.1$ counts~s$^{-1}$ for the secondary source. Figure~\ref{fig_3} shows the model fitted data along with its residual. \par

\begin{table}[htb]
\tabularfont
\caption{Best-fit parameters from the NUV radial profile analysis for the secondary source}
\centering
\begin{tabular}{ lcc }
\hline
Param &Type& Best Value \\
\hline
m1.A  & thawed   &   1145.3 $\pm$ 215.8 \\
m1.$x_0$   & frozen    &   0.6 \\
m1.$\gamma$ & frozen  & 1.7\\
m1.$\beta $ & frozen & 1.6 \\
c1.$c_0$ & thawed  & 21.7 $\pm$29.2 \\
g1.A & thawed &355.5 $\pm$55.7 \\
g1.fwhm &thawed & 3.6 $\pm$ 0.4\\
g1.$x_0$ & thawed & 10.0 $\pm$ 0.1 \\
e1.$x_0$ &thawed& 11.2 $\pm$ 280.3 \\
e1.A & thawed &192  $\pm$ 6847 \\
e1.coeff &thawed & $-0.12 \pm$ 0.03\\
g2.fwhm & thawed & 7.9  $\pm$ 2.7\\
g2.A & thawed & 95.3 $\pm$ 52.8\\
g2.$x_0$ & thawed &  11.4 $\pm$ 1.8 \\ \hline
CPS & & 0.7$\pm$0.1 \\
\hline
\end{tabular}
\label{tab_3}
\end{table}

\vspace{2mm}

Table~\ref{tab_12} summarizes the results obtained from radial fitting.
\begin{table*}[htb]
\tabularfont
\caption{Results from Radial fitting}
\centering
\begin{tabular}{ cccc }
\hline
            Source & Systematic error & $\chi^2$/dof & CPS  \\
            \hline
             Primary FUV & $3.5\%$ &1.0589 & 0.23$\pm$0.03\\
             
             Secondary FUV & $7.0\%$ &1.11381 & 0.09$\pm$0.02\\
             
             Primary NUV& $2.0\%$ & 1.07013&3.9$\pm$0.4\\
        
             Secondary NUV&$2.0\%$ & 1.0872 &0.7 $\pm$0.1\\
             \hline
        
\end{tabular}
\label{tab_12}
\end{table*}

\begin{table*}
    \centering
      \caption{NUV-to-FUV count ratios}
    \begin{tabular}{cccc} \hline
     Source    &  Observed count-rate ratio &  Internal E(B-V) & Predicted  ratio  \\ \hline
     Primary    & $17\pm4$  & 0.36 &  18.2 \\
     Secondary & $8\pm3$    & 0.0 &  7.8\\ \hline
    \end{tabular}
    \label{tab:ratios}
\end{table*}



\subsection{Count-rate ratio analysis} 
The nature of any source can be inferred from its spectrum. In the absence of spectrum, the colors of an object become useful. We calculate the ratio of count rates in the NUV/Silica and FUV/BaF2 bands for the AGN and  the secondary source. These ratios are equivalent to colours. Generally two colors are used to classify or infer the type of an object. In addition to the UV colors, we use X-ray/optical flux.
In Table~\ref{tab:ratios}, we list the UV colours of the two sources. 
As described earlier, we will compare the ratios listed in Table~\ref{tab:ratios} with the corresponding ratios obtained from standard spectra of quasars and starburst galaxies.

\section{Nature of the secondary source}
\subsection{Comparison with composite quasar spectrum}
We used the composite quasar spectrum derived by \cite{c17} using SDSS spectra of over 2200 quasars in the redshift range from 0.044 to 4.789. Since this spectrum has practically zero extinction, but our observations were reddened by both the Galactic extinction and internal reddening in Mrk~766, we need to redden the composite quasar spectrum before calculating the count rates in NUV and FUV.

The 'non-standard' extinction caused by the AGN environment of Mrk~766 was applied with the help of empirical formula given in \cite{c8}. We used a  color excess of E(B-V)=0.36 derived from the Balmer decrement i.e., the ratio of observed H$\alpha$ to H$\beta$ flux of Mrk~766 \citep{c16}. Then we redshifted this reddened composite spectrum for the redshift of Mrk~766 ($z= 0.01293$). Finally we reddened the composite spectrum to account for the Galactic extinction using the CCM89 law \citep{c9}. We assume this reddened and redshifted quasar composite spectrum as our model spectrum for active nucleus in Mrk~766.
We calculated the predicted UVIT count rates for our model spectrum using the effective area of a filter using the following equation:

\begin{equation}
    CPS= \int{\frac{f_{\lambda}}{(hc/\lambda)} A_{eff}}(\lambda)d\lambda
    \label{eqn:eqn2}
\end{equation}

\noindent Using the effective areas for the FUV/BaF2 and NUV/Silica filters, we calculated the predicted count-rate ratio for the composite quasar spectrum to be $18.2$.

We note that the predicted count rate ratio for the composite quasar spectrum is similar to the observed ratio for the AGN but it is very different than the observed ratio for the secondary source. The internal extinction with $\text{E(B-V)}=0.36$ derived from the Balmer decrement is appropriate for the AGN in Mrk~766. It is possible that the secondary source suffers with a different internal extinction.  If we do not redden the composite quasar spectrum with the internal extinction, we predict a count-rate ratio of $7.8$. We list the predicted count-rate ratios in Table~\ref{tab:ratios}.

From Table~\ref{tab:ratios}, we find that the count-rate ratio of the primary source is within the predicted ratio  for the composite quasar spectrum. Thus, our analysis implies that the primary source is indeed an AGN  which in turn verifies the correctness of our methodology. The observed count-rate ratio of the secondary source  deviates significantly from the expected value for an AGN with similar internal reddening as the primary source. But interestingly, the observed ratio for the secondary source matches well with the AGN ratio if there is no internal reddening. Thus, our analysis clearly rules out the secondary source to be a background AGN or an accreting SMBH embedded in the galaxy Mrk~766.  

\subsection{Estimation of X-ray flux and detectability with Chandra}\label{sec_3.2.4}
From our analysis in previous sections, it is clear that the count-rate ratio of the secondary source is consistent with the expected ratio for an unabsorbed AGN. This possibility can be tested by estimating the expected count rate in the X-ray band and comparing it with the upper limit from the \chandra{} data. In order to predict the expected X-ray flux, we use the optical to X-ray flux
ratio $\alpha_{ox}$ which is the ratio of flux densities at 2500\AA\ and $2\kev$.
We first calculate the 2500\AA\ flux density using the observed count rates in the FUV and NUV bands.

We converted the FUV and NUV count rates to flux densities at the mean wavelengths of the filter bandpass using the relation \citep{Tandon_2017}:
\begin{equation}\label{eq01}
    f_{\lambda}(ergs~cm^{-2}~s^{-1}~\textup{\AA}^{-1})=CPS\times UC 
\end{equation}
The unit conversion factor (UC) was derived from the zero point magnitude (ZP) given in \cite{Tandon_2020} and using the relation \citep{Tandon_2017}:
\begin{equation}\label{eq02}
\begin{split}
    \text{ZP}=-2.5 \log_{10}(UC \times \lambda_m^2)-2.407
\end{split}
\end{equation}
Where, $\lambda_m$ is the mean wavelength of the filter in \textup{\AA}.
With $\lambda_m= 2418$\AA\ and $ZP=19.763$, the unit conversion factor for N242W is
\begin{equation}
    UC=2.318 \times 10^{-16}\ (ergs\ s^{-1}\ cm^{-2}\ \textup{\AA}^{-1})/(counts\ s^{-1})
\end{equation}
The observed count rate of $0.7{\rm~counts~s^{-1}}$ for the secondary source is converted to $f_{\lambda}$(2418\AA)$=1.6\times10^{-16}{\rm~ergs~cm^{-2}~s^{-1}~\AA^{-1}}$.

Since we have the observed flux at 2418 \textup{\AA}, we de-reddened it first from the extinction $A_\lambda$ at 2418 \textup{\AA} due to Milky Way using CCM89. Then we simply transferred the wavelength and the corresponding flux to the rest frame of Mrk~766 by multiplying the flux by (1+z) (z=0.01293) and dividing the wavelength by (1+z) which gave us $\lambda=2387$ \textup{\AA}. In order to estimate the flux density at $2500{\rm~\AA}$, we used the composite quasar spectrum.
We scaled the composite quasar spectrum to have the same flux as the secondary source at 2387 \textup{\AA}. With this scaled spectrum, we found $f_{\lambda}(2500\ \textup{\AA})=1.7\times 10^{-16}\ ergs/cm^2/s/\textup{\AA}$.

The optical to X-ray index is given by~\citep{c15},
\begin{equation}
    \alpha_{ox}=-\frac{\log_{10}[(f_{2keV}/f_{2500\textup{\AA}})]}{2.605}
\end{equation}
where $L_{2keV}$ and $L_{2500\textup{\AA}}$ are in ${\rm~ergs~cm^{-2}~s^{-1}~Hz^{-1}}$.

We converted $f_\lambda (2500 \textup{\AA})$ to $f_\nu (2500\textup{\AA})$ and using $\alpha_{ox}=1.37$ \citep{c14} to find $f_\nu(2\kev)=9.76\times 10^{-32}{\rm~ergs~cm^{-2}~s^{-1}~Hz^{-1}}$ or $f_E=7.369\times 10^{-6}$ photons/cm$^2$/s/keV. Using a power law model with X-ray photon index $\Gamma=1.9$ modified with the Galactic absorption $N_H=1.8\times10^{20}{\rm~cm^{-2}}$ along the line of sight to Mrk~766, we calculated $0.4-10\kev$ band X-ray flux, $f_X=1.5\times10^{-13}{\rm~ergs~cm^{-2}~s^{-1}}$.  We converted this flux to \chandra{}/HETG count rate of $0.0017$ counts~s$^{-1}$ using the WebPIMMS tool. This is the expected count rate if the secondary source were an unabsorbed AGN.

We used the \chandra{} observation (ObsID:1597) with an exposure time of 89~ks. There is no X-ray source at the location of the secondary source. We calculated the $4\sigma$ upper limit of 128 counts or 0.0014 counts~s$^{-1}$, which is less than the predicted count rate. Thus, due to the lack of X-ray emission, the secondary is highly unlikely to be another accreting SMBH in Mrk~766. 


%

\begin{figure}
    \centering
    \includegraphics[scale=0.3]{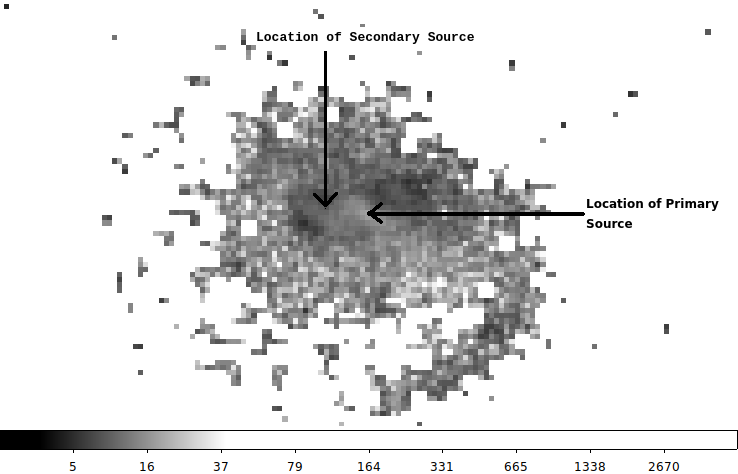}
    \caption{Ratio image obtained by dividing the NUV image by the FUV image after the NUV image was scaled to have the same exposure time as that of the FUV image.}
    \label{fig_8}
\end{figure}

Another possibility is that the secondary source could be a compact region of enhanced star formation. We suspected this possibility based on the ratio image where the secondary source does not stand out. We created a ratio image, shown in Figure~\ref{fig_8}, by dividing each pixel value in the NUV image by the corresponding pixel value in the FUV image after scaling the NUV image to have the same exposure time as the FUV image. We find similar ratios at the position of the secondary source as in other parts of the galaxy except at the location of the primary. This suggests that the emission process responsible for the secondary source is likely similar to that of the diffuse emission from the other parts of the galaxy which is likely from a population of young, massive stars resulting from star formation. 

\subsection{Comparison with the spectra of starburst galaxies}
To investigate further if the secondary source is actually an enhanced star-forming region, we estimated typical NUV-to-FUV count-rate ratios for starburst galaxies.
We obtained the template spectra of starburst galaxies from \cite{c10} derived for different values of the internal extinction, E(B-V). These spectra are already corrected for Galactic extinction.  We calculated the count-rate ratios for the template starburst spectra using Eqn.~\ref{eqn:eqn2} by using the flux densities at the mean wavelengths of our FUV, NUV filters. We did not apply any additional internal reddening. The predicted ratios are listed in Table~\ref{tab_9}. We see that the observed count-rate ratio for the secondary source is very similar to that derived for the starburst templates with internal extinction $E(B-V) < 0.21$. If the secondary source is indeed a compact star forming region with optical/UV spectrum similar to the starburst templates, we expect significant optical emission. 

\begin{figure}[h]
    \centering
    \includegraphics[scale=0.6]{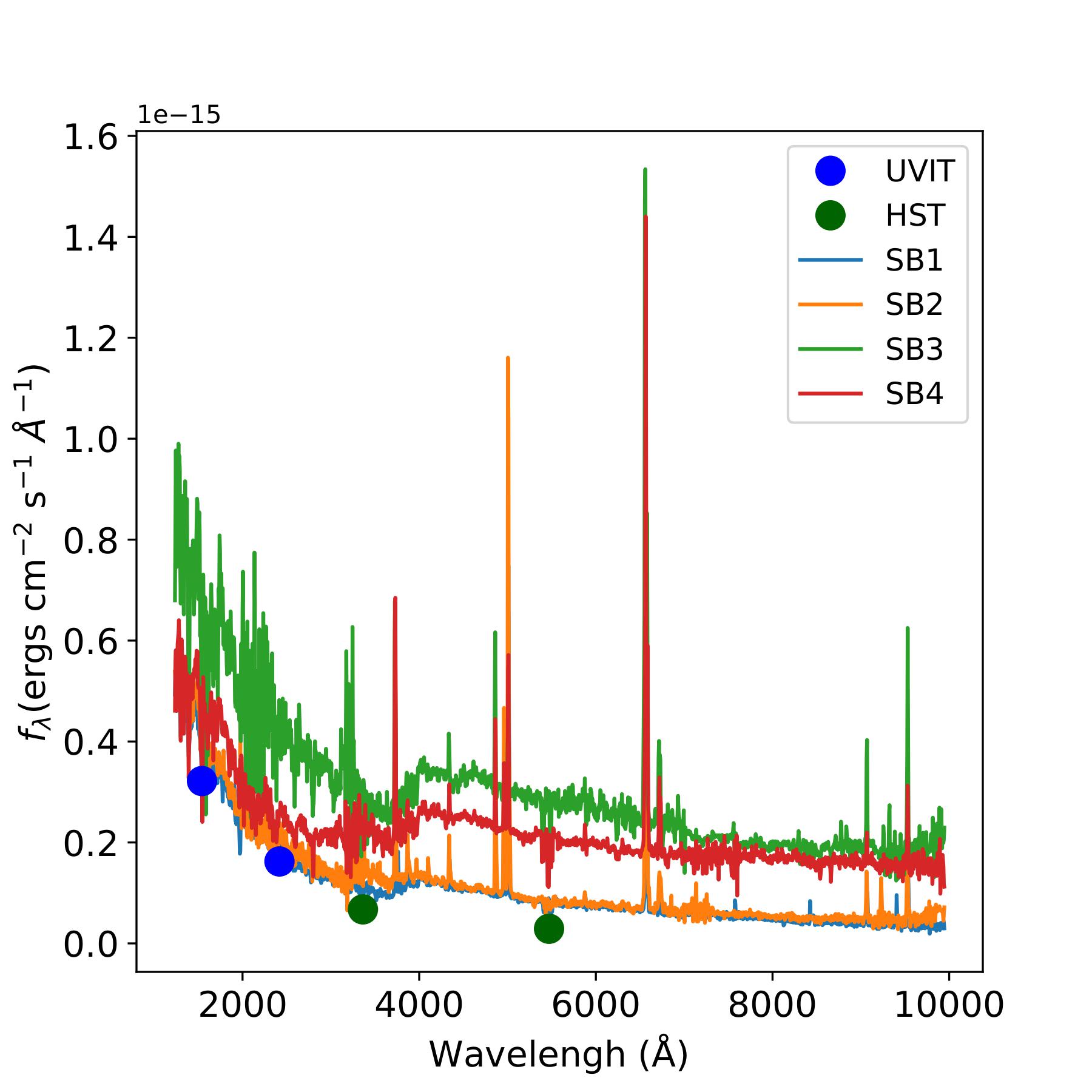}
    \caption{Spectra of starburst galaxies scaled to match the flux density measured with the FUV/F154W filter at the mean wavelength $1541\AA$. Filled circles show flux densities  measured with the NUV/N242W filter ($\lambda_{mean}=2418\AA$), HST/ACS F330W filter ($\lambda_{central}=3362.7\AA$) and HST/WFC3 F547M filter ($\lambda_{central}=5475\AA$) are also shown.}
    \label{fig10}
\end{figure}

    \begin{table}[]
        \centering
        \tabularfont
        \caption{Ratio obtained from UV templates of starburst galaxies}
        \begin{tabular}{lcc}
        \hline
            Template & E(B-V) & Ratio \\
            \hline
             Starburst 1&  $E(B-V)< 0.1$ & \textbf{7.36}\\
             
             Starburst 2& $0.11 < E(B-V)<0.21$ & \textbf{7.5}\\
             
             Starburst 3& $0.25< E(B-V)<0.35$ & \textbf{10.1}\\
        
             Starburst 4& $0.39< E(B-V)<0.50$ & \textbf{8.3}\\
             \hline
        \end{tabular}
        \label{tab_9}
    \end{table}

\subsection{HST images and calculation of flux}
\begin{figure}
     \centering
     \includegraphics[scale=0.3]{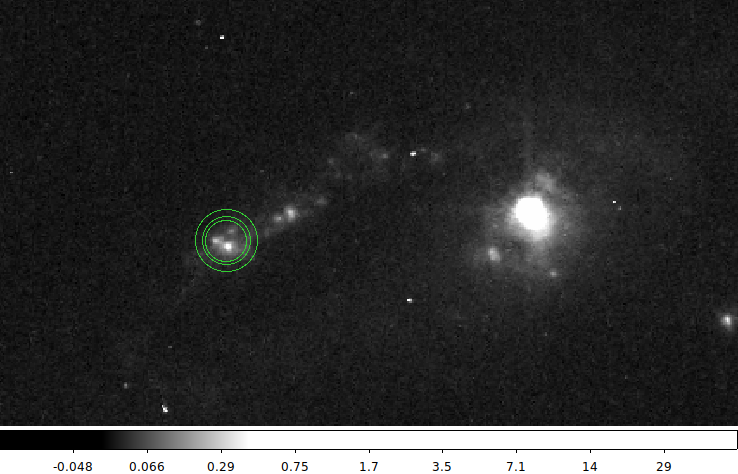}
     \caption{Selected regions for aperture photometry on the secondary source. }
     \label{fig_ap}
 \end{figure}

We searched for the counterpart of the secondary source in \hst{} images.
We detected multiple compact sources near the position of the secondary source in the \hst{} images of Mrk~766 acquired with different instrument and filter combinations. Because of the excellent PSF of the \hst{}, the secondary source that appeared as a point source in UVIT images, now appeared as three distinguishable point sources. Figure~\ref{fig_ap} shows the secondary source marked with circles in the \hst{} image in  the F330W filter. We included the three sources and performed aperture photometry and obtained the count rates for the secondary source in  the \hst{}/ACS F330W filter (central wavelength $=3362.7~\AA$) and \hst{}/WFC3 F547M filter (central wavelength $=5475~\AA$). We used a circular region of radius $0.3$ arc-seconds to encircle the source as shown in Figure \ref{fig_ap} and another concentric annular region of inner radius 0.35 arc-seconds and outer radius 0.45 arc-seconds to  estimate the background count rate. We then calculated the background corrected net count rates and converted to flux densities using the  value of the PHOTFLAM keyword in the headers of the \hst{} images. We found $f_{\lambda}$(3363\AA)=$6.74\times10^{-17}{\rm~ergs~cm^{-2}~s^{-1}}$ for F330W filter and  $f_{\lambda}$(5475\AA)=$2.9\times10^{-17}{\rm~ergs~cm^{-2}~s^{-1}}$ for the F547M filter.

In order to compare these flux densities measured with the HST  and that expected from the secondary source assuming it to be a compact star forming region, we re-scaled the starburst template spectra to the measured flux density at $2418~\AA$ (shown in section \ref{sec_3.2.4}) with the UVIT/NUV. We then compared the UV and optical flux densities with the re-scaled starburst template spectra in Figure~\ref{fig10}. We find that the measured flux at four different wavelengths follow the shape of the starburst template spectra. This clearly suggests that the secondary source is indeed a compact star forming region. 

\vspace{-2em}
\section{Conclusion} \label{conclusion}
We identified a bright far and near UV source at a projected distance of $\sim 1.1{\rm~kpc}$ from the known active nucleus. Such pairs of compact sources can easily be suspected as a pair of accreting SMBHs. We investigated the nature of the secondary source using NUV-to-FUV flux ratio, \chandra{} X-ray observation, and \hst{} images in the near UV and optical bands. The lack of X-ray emission in the \chandra{} image at the location of the secondary source makes it highly unlikely to be an accreting SMBH. Further, the UV/optical flux measured at four different bands closely follow the shape of the starburst template spectra. Therefore we conclude that the secondary is most likely a compact star forming region. 







\section*{Acknowledgements}
This publication uses the UVIT data from the AstroSat mission of the Indian Space
Research Organisation (ISRO), archived at the Indian Space Science Data Centre
(ISSDC). The UVIT data are processed by the payload operations centers at IIA, Bangalore. This research has made use of UVIT pipeline (CCDLAB) developed at  University of Calgary for UVIT development and science support. The scientific results reported in this article are based in part  on observations made by the Chandra X-ray Observatory, data obtained from the Chandra Data Archive. This research is based on observations made with the NASA/ESA Hubble Space Telescope obtained from the Space Telescope Science Institute, which is operated by the Association of Universities for Research in Astronomy, Inc., under NASA contract NAS 5–26555.
This research has made use of the $python$ and $julia$ packages.
This research has made use of the SIMBAD database, operated at CDS,
Strasbourg, France. 

\vspace{-1em}








\bibliography{bibfile}

\end{document}